\documentclass[a4paper,10pt,twoside]{cpc-hepnp}

\usepackage{multicol}
\usepackage{graphicx}
\usepackage{booktabs}
\usepackage{amssymb,bm,mathrsfs,bbm,amscd}
\usepackage[tbtags]{amsmath}
\usepackage{lastpage}

\begin{document}

\fancyhead[co]{\footnotesize YUAN Yao-Shuo et al: The study of beam loading effect in the CSNS/RCS}


\title{The study of beam loading effect in the CSNS/RCS\thanks{Supported by National Natural Science
Foundation of China (11175193) }}

\author{%
  YUAN Yao-Shuo(ԷҢ˶)$^{1}$
  \quad LI Kai-Wei(Àçâ)$^{1}$
      \quad WANG Na(ÍõÄÈ)$^{1}$
  \quad Yoshiro Irie$^{2}$
\quad WANG Sheng(ÍõÉú)$^{1;1)}$ \email{wangs@ihep.ac.cn. Corresponding author.}%
}
\maketitle

\address{%
$^1$ Institute of High Energy Physics, Chinese Academy of Sciences, Beijing 100049, China\\
$^2$ KEK, High Energy Accelerator Research Organization, 1-1 Oho, Tsukuba-shi, Ibaraki-ken 305-0801, Japan\\
}

\begin{abstract}
CSNS/RCS accelerates a high-intensity proton beam from 80 MeV to 1.6 GeV. Since the beam current and beam power is high, the beam loading is a severe problem for the stability of the circulating beam in the RCS. To study the beam loading effect in the CSNS/RCS theoretically, the RLC circuit model of the rf cavity, the method of the Fast Fourier Transform and the method of Laplace transform have been employed to obtain the impedance of the rf system, the beam spectrum and the beam-induced voltage, respectively. Based on these physical models, the beam dynamics equations have been revised and a beam loading model has been constructed in the simulation code ORIENT. By using the code, the beam loading effect on the rf system of the CSNS/RCS has been investigated. Some simulation results have been obtained and conclusions have been drawn.
\end{abstract}



\begin{multicols}{2}

\section{Introduction}

Rapid Cycling Synchrotron (RCS) is a key component of the Chinese Spallation Neutron Source (CSNS)\cite{Wang1}. The kinetic energy of the beam reaches 80 MeV after accelerated through the linac. After the beam is injected into the RCS, the beam kinetic energy and the beam power can reach 1.6GeV and 100 kW respectively.

For this kind of synchrotron, how to control the beam loss to a low level is an important issue. According to the operating experience in other existing machines in the world, beam loading effect is one of the most important contributions to the beam loss \cite{Ezura1}\cite{Garoby1}\cite{Barratt1}. It affects both the amplitude and the relative phase of the voltage seen by the beam as it passes through the rf cavity. Specifically, as the beam passes through the cavity, it can induce wakefield and the corresponding voltage will affect the distribution of the electromagnetic field and voltage in the cavity. As a result, the effective voltage, which is a superposition of rf voltage produced by the generator and the beam-induced voltage, will influence the beam longitudinal motion and make the instability and even the beam loss happen.

A new simulation code, named ORIENT\cite{Yuan1}, has been developed for the optimization of the rf voltage waveform and simulation of the longitudinal beam motion in the CSNS/RCS. With the beam loading model developed in the code, it can be used for simulating and evaluating the beam loading effect in the CSNS/RCS. The work has been performed in two steps: firstly, the impedance of the rf cavity can be calculated using the equivalent RLC circuit of the cavity and the beam spectrum can be obtained by using the fast Fourier transformation (FFT); secondly, the beam-induced voltage can be calculated with the method of Laplace transform and the influence of the beam-induced voltage on the beam itself, such as the beam distribution in the phase space, the beam loss and the ratio of the beam-induced voltage to the effective voltage can be calculated by using revised particle tracking equations with beam loading effect considered.

\section{The equivalent RLC circuit model of the rf cavity}

The rf cavity employed in the CSNS/RCS is a low frequency ferrite loaded coaxial cavity. The main parameters are listed in Table~\ref{tab1}\cite{CSNS1}. From the view of the circuit characteristics, the rf cavity in the CSNS/RCS can be equivalent to two RLC parallel resonance circuits, since the cavity includes two accelerating gaps. The equivalent circuit is shown in Fig.~\ref{Fig.1.}. Here the generator and the beam are viewed as two current sources and $C_{1}$£¬$L_{1}$£¬$R_{1}$ and $C_{2}$£¬$L_{2}$£¬$R_{2}$ represent the value of the capacitances of the accelerating gap, the equivalent inductance and the shunt impedance in each parts of the cavity, respectively. $I_{g}$ and $I_{b}$ denote the current of the generator and the beam.

\begin{center}
\tabcaption{ \label{tab1}  Main parameters of the ferrite-loaded cavity in the CSNS/RCS}
\footnotesize
\begin{tabular*}{80mm}{c@{\extracolsep{\fill}}cc}
\toprule Parameters/units & Values   \\
\hline
Frequency/MHz & 1.02$\sim$2.44  \\
Total rf voltage/kV & 165 \\
Cavity numbers & 8 \\
Cavity length/m & 2.71\\
Number of gaps & 2 \\
Gap voltage/kV & 10.3 \\
Capacitance per gap/nF  & 3\\
Inductance per gap/$\mu$H   &  8.1$\sim$1.4\\
\bottomrule
\end{tabular*}
\end{center}

\begin{center}
\includegraphics[width=7cm]{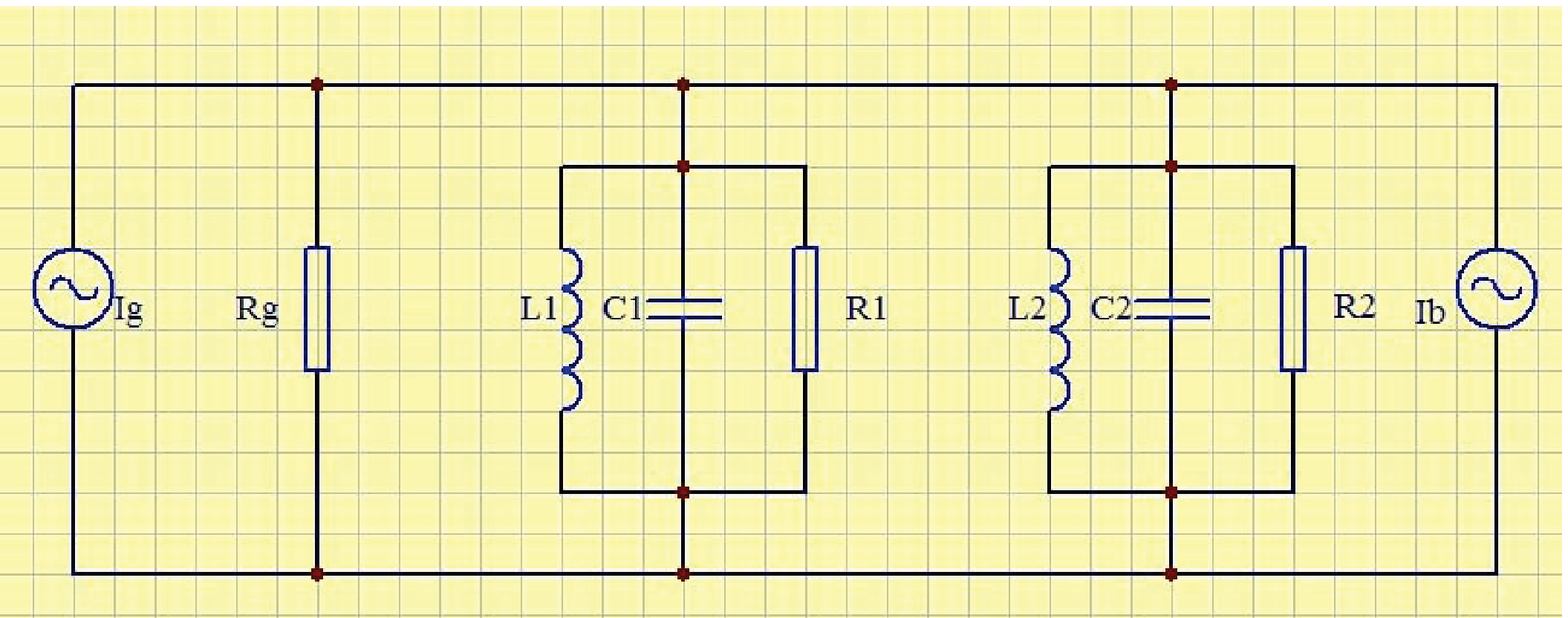}
\figcaption{\label{Fig.1.} Schematic drawing of the direct-equivalent RLC circuit of the rf cavity}
\end{center}

\begin{center}
\includegraphics[width=7cm]{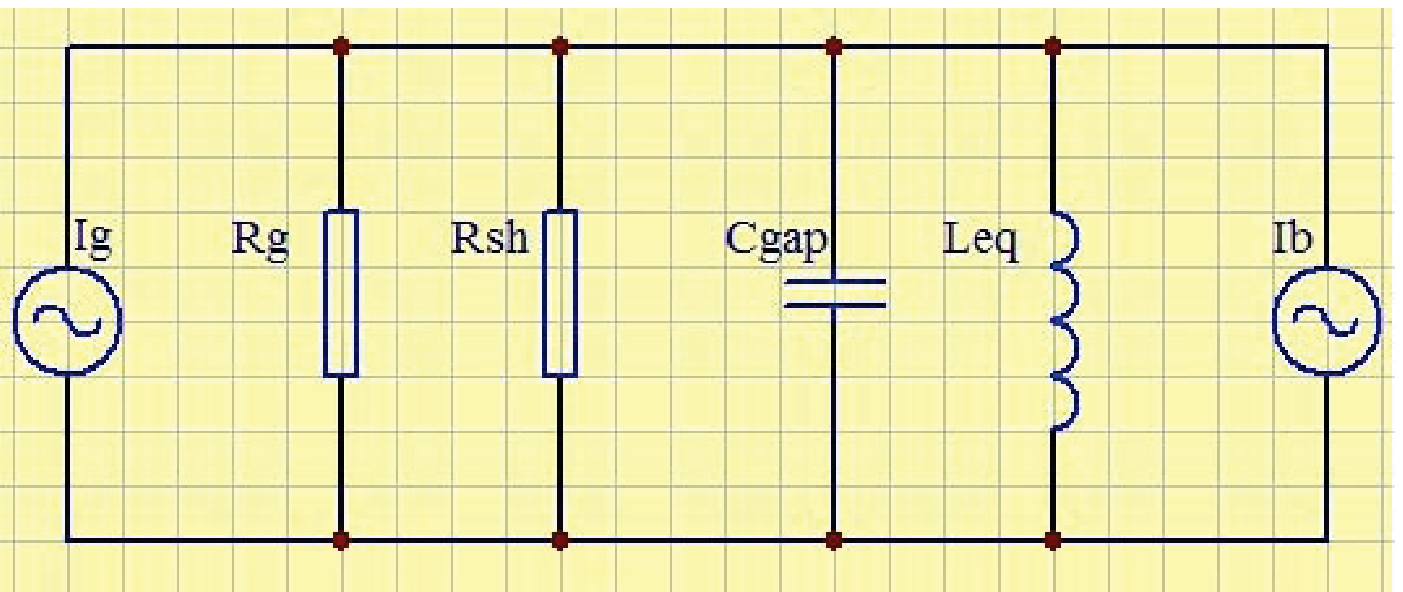}
\figcaption{\label{Fig.2.} Schematic drawing of the simplified equivalent RLC circuit of the rf cavity}
\end{center}

The equivalent circuit in Fig.~\ref{Fig.1.}, which contains a combination of two RLC circuit, can be further simplified to one RLC circuit shown in Fig.~\ref{Fig.2.}\cite{Pedersen1}, the corresponding parameters are given by

\begin{equation}
\label{eq1}
\left\{
\begin{aligned}
R_{sh}&=& R_{1}+R_{2}\\
C_{gap}&=&C_{1}+C_{2}\\
L_{eq}&=&L_{1}+L_{2}
\end{aligned}
\right.
\end{equation}

The total impedance of the RLC circuit $Z_{sum}$, which can represent the impedance of the rf cavity, can be written as

\begin{eqnarray}
\label{eq2}
Z_{sum}(\omega)=( \frac{1}R+\frac{1}{j{\omega}L_{eq}} + j{\omega}C_{gap})^{-1}
\end{eqnarray}
where  R=$R_{g}$$R_{sh}$/($R_{g}$+$R_{sh}$).

\section{The beam-induced voltage}

According to the Kirchhoff's laws, the current equation for the circuit in Fig.~\ref{Fig.2.} can be written as,

\begin{equation}\label{eq3}
    i_{Rg}+i_{Rsh}+i_{L}+i_{C}=i_{g}+i_{b},
\end{equation}
where the subscript means the current in the branch of that element. The generator current can be written as,

\begin{equation}\label{eq4}
    i_{g}=I_{g}cos(\omega_{r}t+\phi_{g})
\end{equation}
where $\omega_{r}$ is the resonance angular frequency and $\phi_{g}$ is the initial phase of the generator. The current of the beam circulating in the ring can be expanded as a Fourier series with the resonance frequency of the rf cavity as the fundamental frequency,

\begin{equation}\label{eq5}
    i_{b}=\sum_{k=0}I_{bk}cos(k\omega_{r}t+\phi_{g})
\end{equation}
where $I_{bk}$, $\phi_{bk}$ denotes the amplitude and the phase of the $k^{th}$ order respectively. In fact, the quality factor $Q$ of the ferrite-loaded cavity in the CSNS/RCS may reach up to 100 in an acceleration period. Therefore the first-order component ($k$=1) in Eq.~\ref{eq5}, has the largest contribution to the beam-induced voltage in the cavity and the contribution made by other harmonic components can be neglected. Eq.~\ref{eq5} can be simplified to

\begin{equation}\label{eq6}
    i_{b}=I_{b1}cos(\omega_{r}t+\phi_{b1}).
\end{equation}

For calculating the beam-induced voltage, the Laplace transform method can be employed. After transformed from the time domain to the s-domain (i.e. the Laplace domain), the whole impedance and the beam current in Eq.(~\ref{eq2}) and Eq.(~\ref{eq6}) become

\begin{equation}
\label{eq7}
\widetilde{Z}_{sum}(s)=\frac{1}{C_{gap}}\frac{s}{s^2+s/C_{gap}R+1/L_{eq}C_{gap}}, \\
\end{equation}

\begin{equation}\label{eq8}
\widetilde{i_{b}}=\frac{s^2+w_{r}^2}{scos\phi_{b1}-\omega_{r}sin\phi_{b1}}
\end{equation}
where $s$ is the independent variable after Laplace transform.

The expression of beam-induced voltage in s-domain can be calculated from the above two equations,

\begin{equation}
\label{eq9}
\widetilde{v}_b=\widetilde{i}_b\widetilde{Z}_sum=\frac{1}{C_{gap}}\frac{s}{s^2+s/(C_{gap}R)+1/(L_{eq}C_{gap})}
\end{equation}

With the beam current as the periodical driving source, the beam-induced voltage in the cavity would contain a stable term and an attenuation term, therefore it can be written as

\begin{equation}\label{eq10}
    \widetilde{v}_b=\frac{ms+n\omega_{r}}{s^2+\omega_{r}^2}+\frac{ks+l}{s^2+s/(C_{gap}R)+1/(L_{eq}C_{gap})}.
\end{equation}

Combining Eq.(~\ref{eq9}) and Eq.(~\ref{eq10}), and taking the resonance condition $\omega_{r}^2=1/(L_{eq}C_{gap})$  into account, the coefficients $m$, $n$, $k$, $l$ can be solved by using the method of undetermined coefficients,

\begin{equation}
\label{eq11}
\left\{
\begin{aligned}
  m &=& Rcos\phi_{b1} \\
  n &=& -Rsin\phi_{b1} \\
  k &=& m \\
  l &=& \frac{-Rsin\phi_{b1}}{\sqrt{L_{eq}C_{gap}}}
\end{aligned}
\right.
\end{equation}

After performing the inverse Laplace transform on the both terms in Eq.(~\ref{eq10}), we can obtain,

\begin{equation}\label{eq11}
    \pounds^{-1}(\frac{ms+n\omega_{r}}{s^2+\omega_{r}^2})=mcos\omega_{r}t+nsin\omega_{r}t
\end{equation}

\begin{equation}
\begin{aligned}
&\pounds^{-1}(\frac{ks+l}{s^2+s/(C_{gap}R)+1/(L_{eq}C_{gap})})\\
&
=(Rcos\phi_{b}cos\omega't+\frac{-sin\phi_{b}R\omega-cos\phi_{b}/(2C)}{\omega'}sin\omega't)e^{-{\xi}t}
\end{aligned}
\end{equation}

where $\omega'$=($\omega^2$-1/(4$C^2$$R^2$))$^{1/2}$, and $\xi$=1/(2RC). There appears an exponential decay term after the inverse Lapalace transform. In fact, the total capacitance of the two accelerating gaps in an rf cavity of CSNS/RCS is 6 nF, and the shunt impedance is about several hundred ohm, which means that the term decreases very quickly, so it is reasonable that only the stable term need to be considered in the evaluation and simulation of the beam loading effect. The beam-induced voltage becomes a concise form under the resonance condition of the circuit:

\begin{equation}\label{eq14}
    v_{b}=RI_{b}cos(\omega_{r}t+\phi_{b1})
\end{equation}

\section{Simulation model with beam loading effect}

\subsection{The phasor diagram of the voltages}

The generator voltage can be written as a sinusoidal form from the generator current in Eq.(~\ref{eq4}) and for simplicity, we assume the initial phase of the generator $\phi_{g}$ equals zero,

\begin{equation}\label{eq15}
    v_{g}=V_{g}sin(h\omega_{r}t)
\end{equation}

Similarly, the beam-induced voltage can be written as a sinusoidal form from Eq.(~\ref{eq14}),

\begin{equation}\label{eq16}
    v_{b}=RI_{b}sin(\omega_{r}t+\phi_{b})
\end{equation}

in which

\begin{equation}\label{eq17}
    \phi_{b}=\phi_{b1}+\frac{\pi}2
\end{equation}

The generator voltage and the beam-induced voltage in the form of Eq.(~\ref{eq15}) and Eq.(~\ref{eq16}) can be expressed as two phasors in a phasor diagram, as shown in Fig.~\ref{Fig.3.}. It can be seen in the figure that both the amplitude and the phase of the effective voltage acted on the synchronous particle and the synchronous phase have changed with beam loading effect considered.

\begin{center}
\includegraphics[width=6cm]{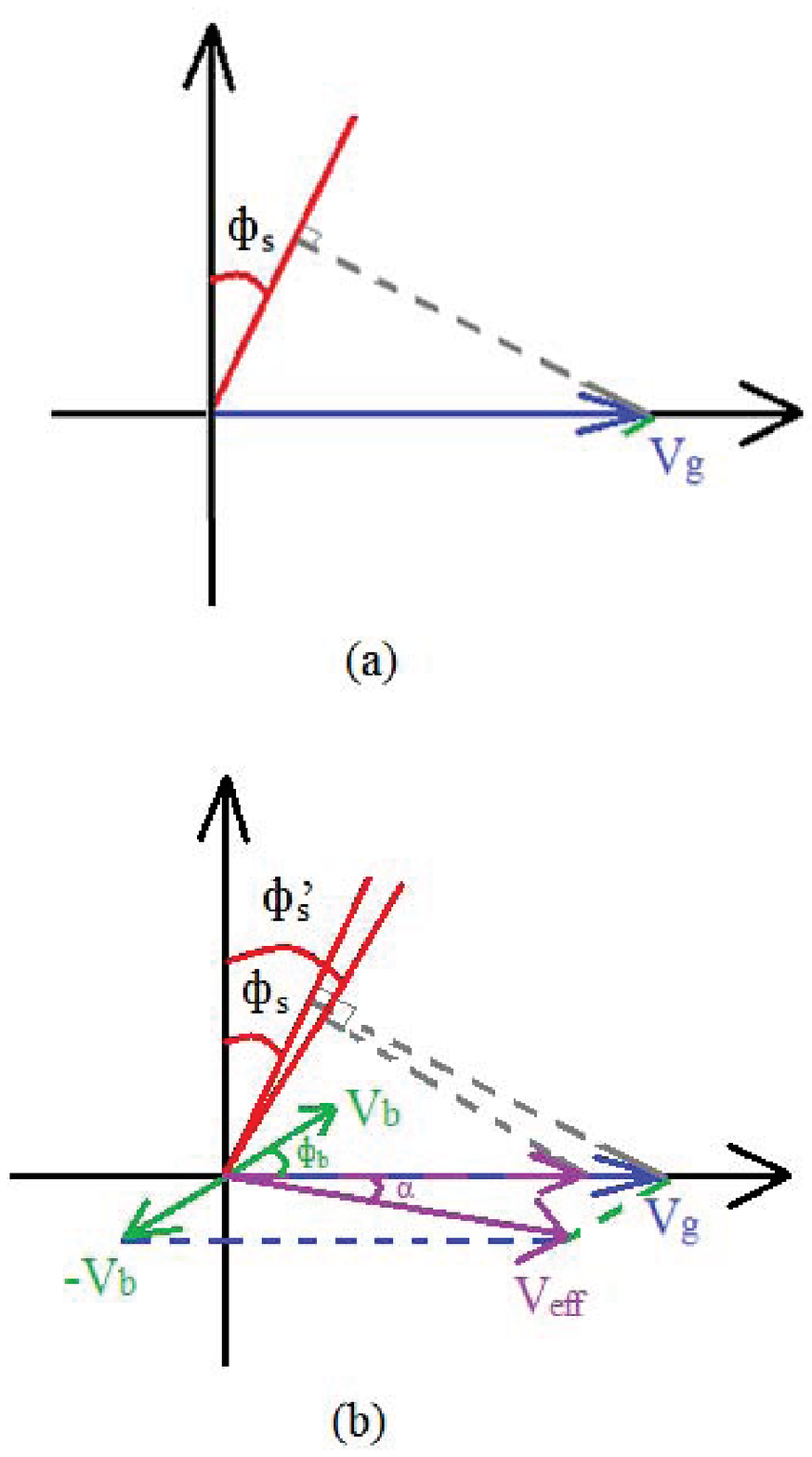}
\figcaption{\label{Fig.3.}  The relation between voltage phasors (a) without and (b) with beam loading effect }
\end{center}

The amplitude and the relative phase of the effective voltage, $V_{eff}$ and $\alpha$ can be written as

\begin{equation}\label{eq18}
    V_{eff}=\sqrt{(V_{g}-V_{b}cos\phi_{b})^2+V_{b}^2sin^2\phi_{b}}
\end{equation}

\begin{equation}\label{eq19}
    \alpha=tan^{-1}(\frac{V_{b}sin\phi_{b}}{V_{g}-V_{b}cos\phi_{b}})
\end{equation}

According to the principle of the phase stability, the relation between the voltage and the synchronous phase is determined by the derivative of the dipole field in the ring with,

\begin{equation}\label{eq20}
    \rho L\frac{dB(t)}{dt}=V_{eff}sin\phi_{s}'
\end{equation}

The new synchronous phase $\phi_{s}'$ can be found from Eq.(~\ref{eq20}) and Eq.(~\ref{eq18}). The synchronous phase becomes larger with the beam loading, for the reason that the beam-induced voltage cancels part of the voltage supplied by the generator (generator voltage) (see Fig.~\ref{Fig.4.}).

\begin{center}
\includegraphics[width=8cm]{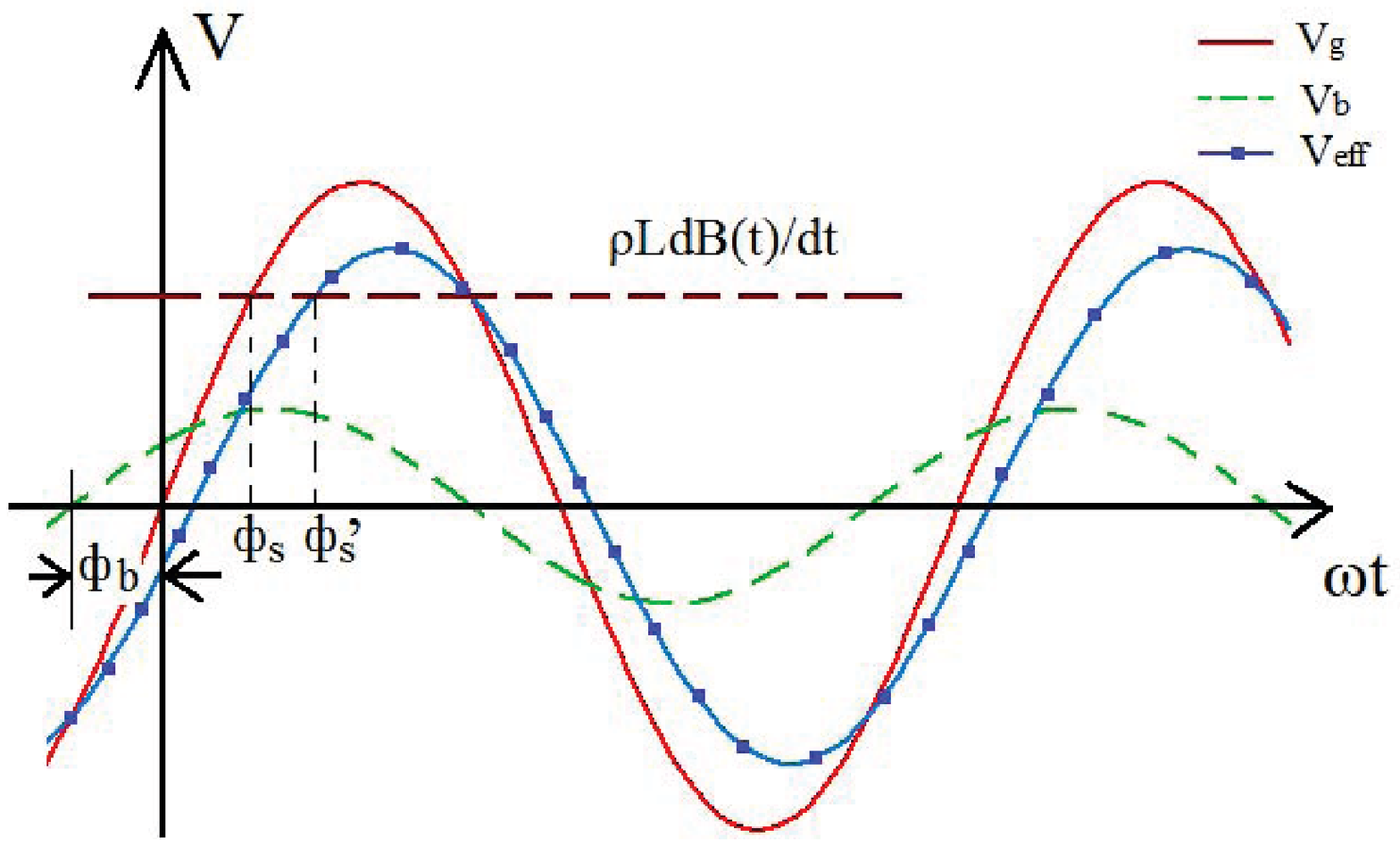}
\figcaption{\label{Fig.4.}  The change of the waveform of the voltage and the synchronous phase with beam loading effect }
\end{center}

\subsection{Beam dynamics equations with beam loading effect}
After the voltages are expressed in the form of phasors, the relation among voltages becomes distinct and the longitudinal particle tracking with beam loading effect can be achieved by revising the beam dynamics equations. With the beam loading effect, both the phase of the synchronous particle and the non-synchronous particles are changed. However, the methods to solve them are different: the former can be obtained according to Eq.(~\ref{eq18}) and Eq.(~\ref{eq20}), while the latter can be found according to the change of the phase of voltage seen by the particle. Let's consider the $i^{th}$ particle. In the absence of the beam loading effect, we assume its phase relative to the generator voltage is $\pi$/2-$\phi_{i}$ and the energy of the particle acquired from the rf cavity can be written as

\begin{equation}\label{eq21}
    \triangle E=eV_{g}sin\phi_{i}
\end{equation}

With the beam loading effect, its phase relative to the effective voltage changes to $\pi$/2-$\phi_{i}$+$\alpha$ , as is shown in Fig.~\ref{Fig.5.}. The energy acquired from the rf cavity would change to,

\begin{equation}\label{eq22}
    \triangle E'=eV_{eff}sin(\phi_{i}-\alpha)
\end{equation}

\begin{center}
\includegraphics[width=6cm]{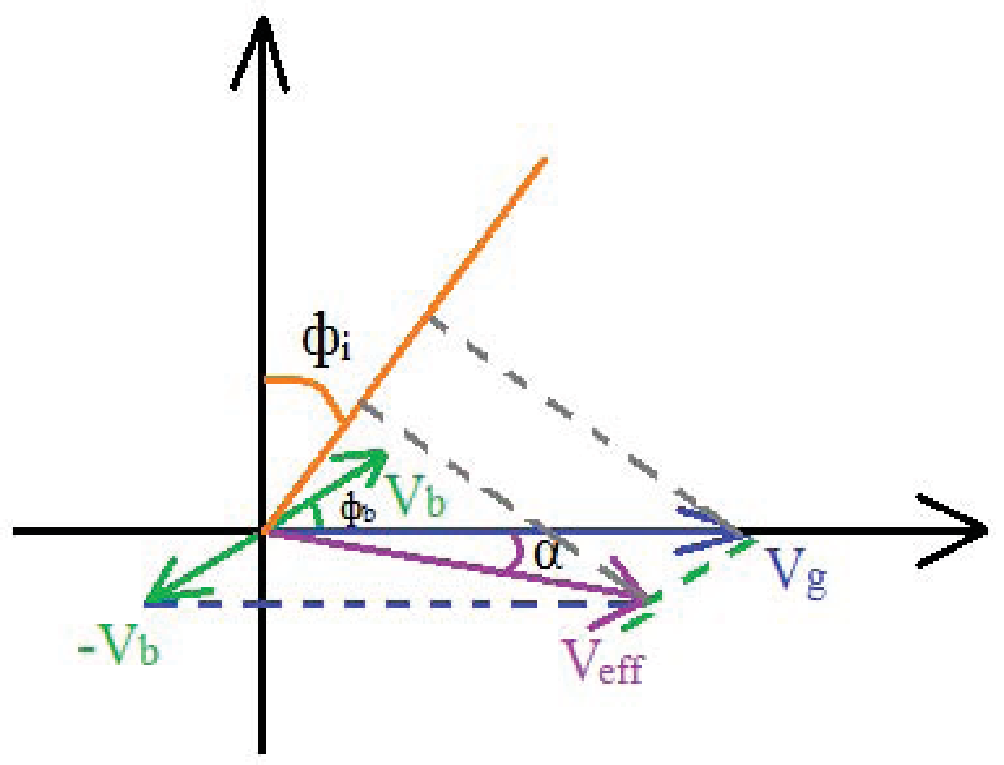}
\figcaption{\label{Fig.5.}  The effective voltage in rf cavity seen by the $i^{th}$ particle }
\end{center}

The longitudinal tracking equation set of the $i^{th}$ particle at $n^{th}$ turn can be revised to,

\begin{equation}\label{eq23}
    \frac{\triangle E_{i}^{(n+1)}}{\omega_{s}^{(n+1)}}=\frac{\triangle E_{i}^{(n)}}{\omega_{s}^{(n)}}+\frac{eV_{total}^{(n)}}{\omega_{s}^{(n)}}[sin(\phi_{i}^{(n)}-\alpha^{(n)})-sin(\phi_{s}'^{(n)})]
\end{equation}

\begin{equation}\label{eq24}
    \phi_{i}^{(n+1)}=\phi_{i}^{(n)}+\frac{2\pi\omega_{s}^{(n+1)}h\eta^{(n+1)}}{(\beta_{s}^{(n+1)})^2E_{s}^{(n+1)}}\frac{\triangle E_{i}^{(n+1)}}{\omega_{s}^{(n+1)}}+\triangle\phi_{s}^{(n)}
\end{equation}
where $\omega_{s}$£¬$\beta_{s}$£¬$E_{s}$ denote the revolution angular frequency, the relativistic speed and the energy of the synchronous phase. $e$, $h$ and $¦Ç$ denotes the unit charge, harmonic number and the slip factor, respectively, and $\triangle\phi_{s}^{(n)}$= $\phi_{s}^{(n+1)}$-$\phi_{s}^{(n)}$.

\section{Simulation study for the CSNS/RCS}
The physical model of beam loading effect has been developed in the code ORIENT based on the theory and equation described above. By using the code, simulation of the beam loading effect in the CSNS/RCS has been performed and several conclusions have been drawn.
The value of the rf voltage and shunt impedance on each millisecond during one acceleration period and the basic parameters of the CSNS/RCS employed as the input data of the code are drawn in Fig.~\ref{Fig.6.} and listed in Table~\ref{tab1}. Fig.~\ref{Fig.7.} to Fig.~\ref{Fig.9.} shows the results of the calculation simulation.

\begin{center}
\includegraphics[width=6cm]{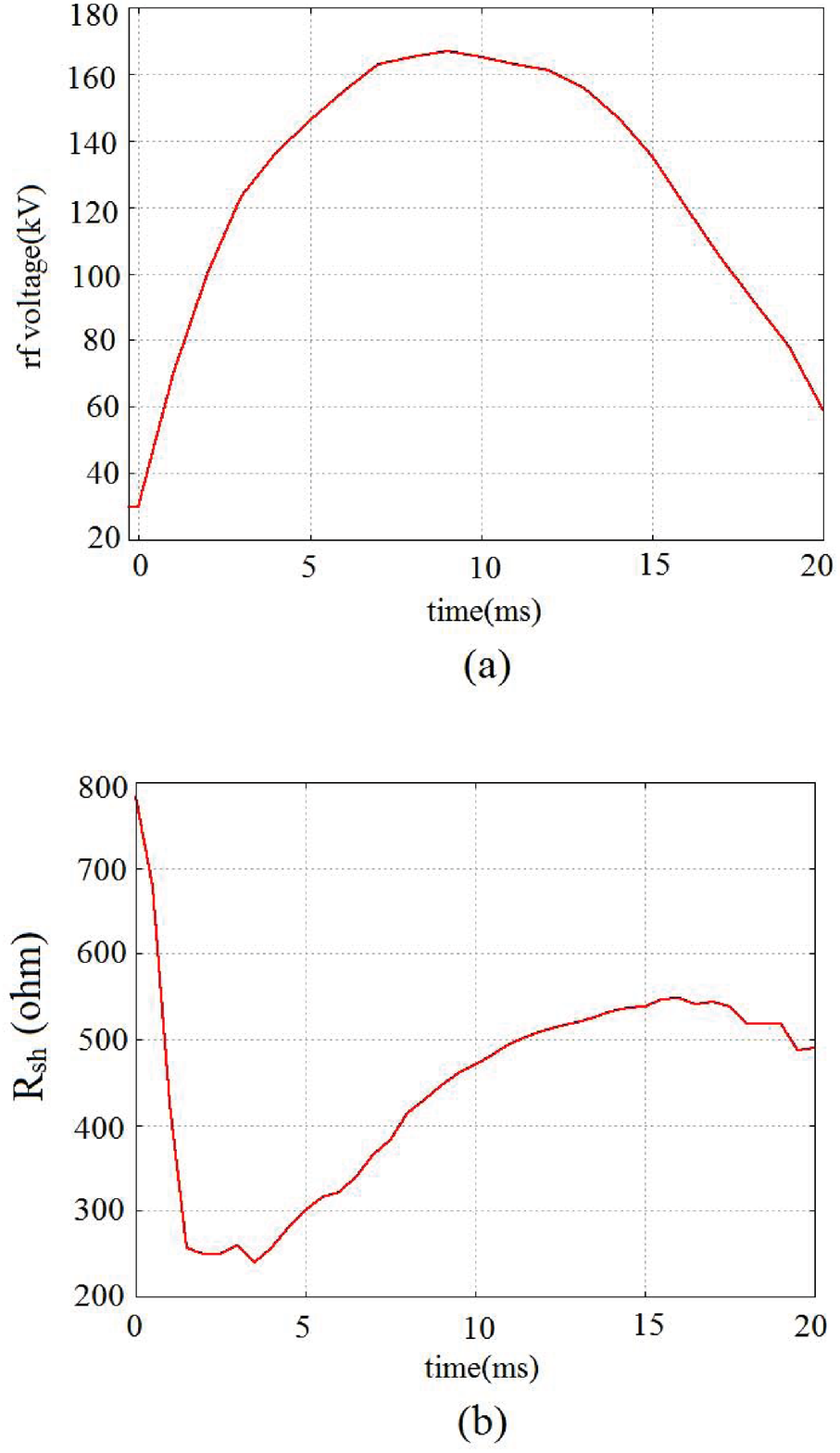}
\figcaption{\label{Fig.6.}  The curve of (a) rf voltage and (b) shunt impedance of rf cavity in an acceleration period }
\end{center}

\begin{center}
\tabcaption{ \label{tab2}  The input parameters of the RCS and the bunch for simulation}
\footnotesize
\begin{tabular*}{80mm}{c@{\extracolsep{\fill}}cc}
\toprule Parameters/units & Values   \\
\hline
Injection starting time (ms) & -0.3  \\
Injection energy(MeV) & 80.0 \\
The number of turns during injection & 200 \\
The numbers of macro particles per turn & 50\\
The number of protons per bunch & 7.8$\times$$10^{12}$ \\
Energy divergence (MeV) & 0.1 \\
Longitudinal distribution of beam & Uniform\\
Energy distribution of beam  &  Gaussian\\
Chopping factor  &   0.5 \\
Harmonic number  &  2  \\
Repetition rate(Hz)  &  25\\
\bottomrule
\end{tabular*}
\end{center}

The changes of the amplitude of the voltage acted on the synchronous particle and the synchronous phase are shown in Fig.~\ref{Fig.7.}. The beam-induced voltage and the ratio of the beam-induced voltage to the generator voltage calculated by the code are shown in Fig.~\ref{Fig.8.}. The comparisons of the particle distribution in the longitudinal phase space without and with beam loading are plotted in Fig.~\ref{Fig.9.} from the code.

\begin{center}
\includegraphics[width=6cm]{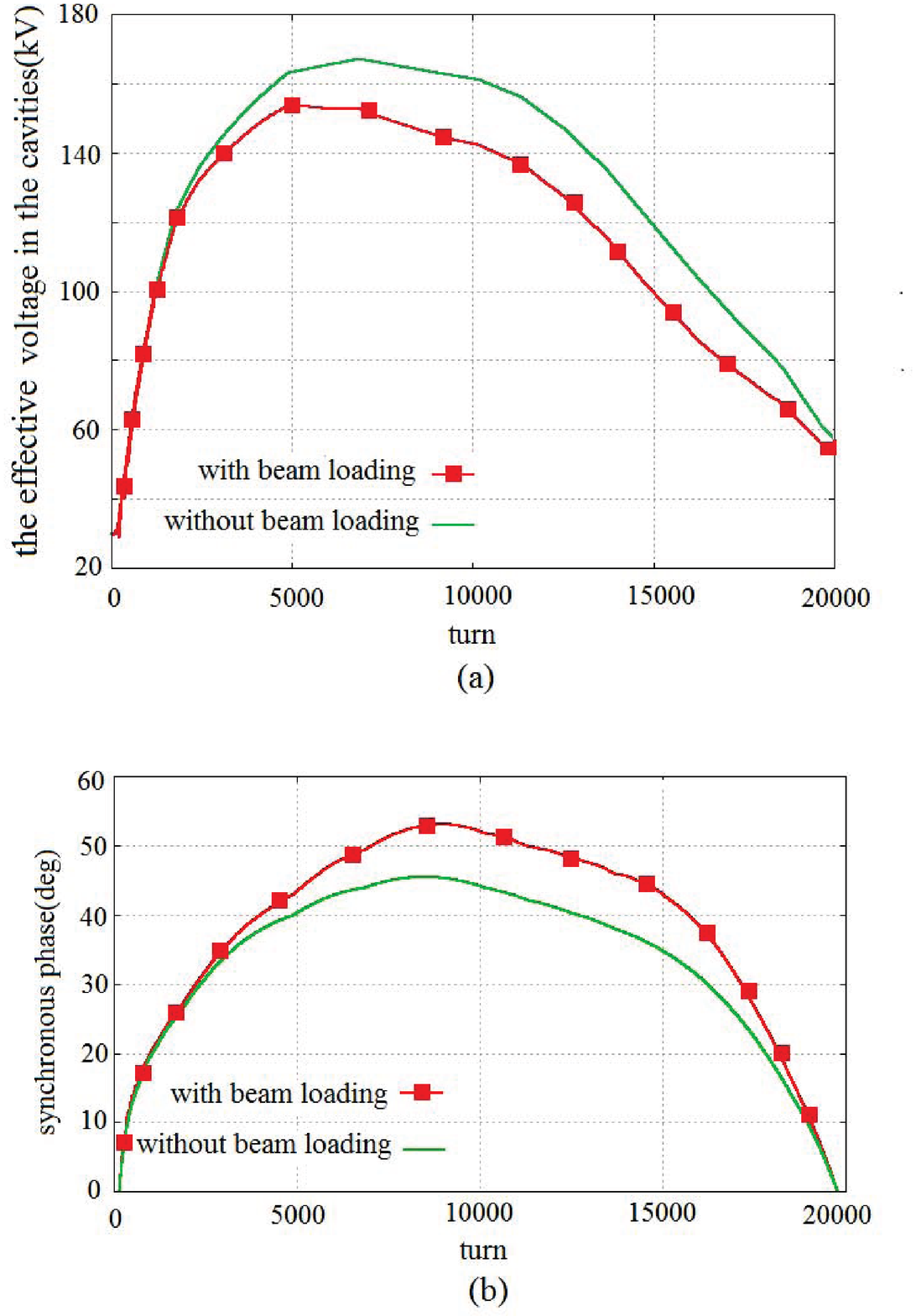}
\figcaption{\label{Fig.7.}  (a) The change of the amplitude of voltage in the rf cavity and (b) the synchronous phase with beam loading effect in one acceleration period }
\end{center}

\begin{center}
\includegraphics[width=6cm]{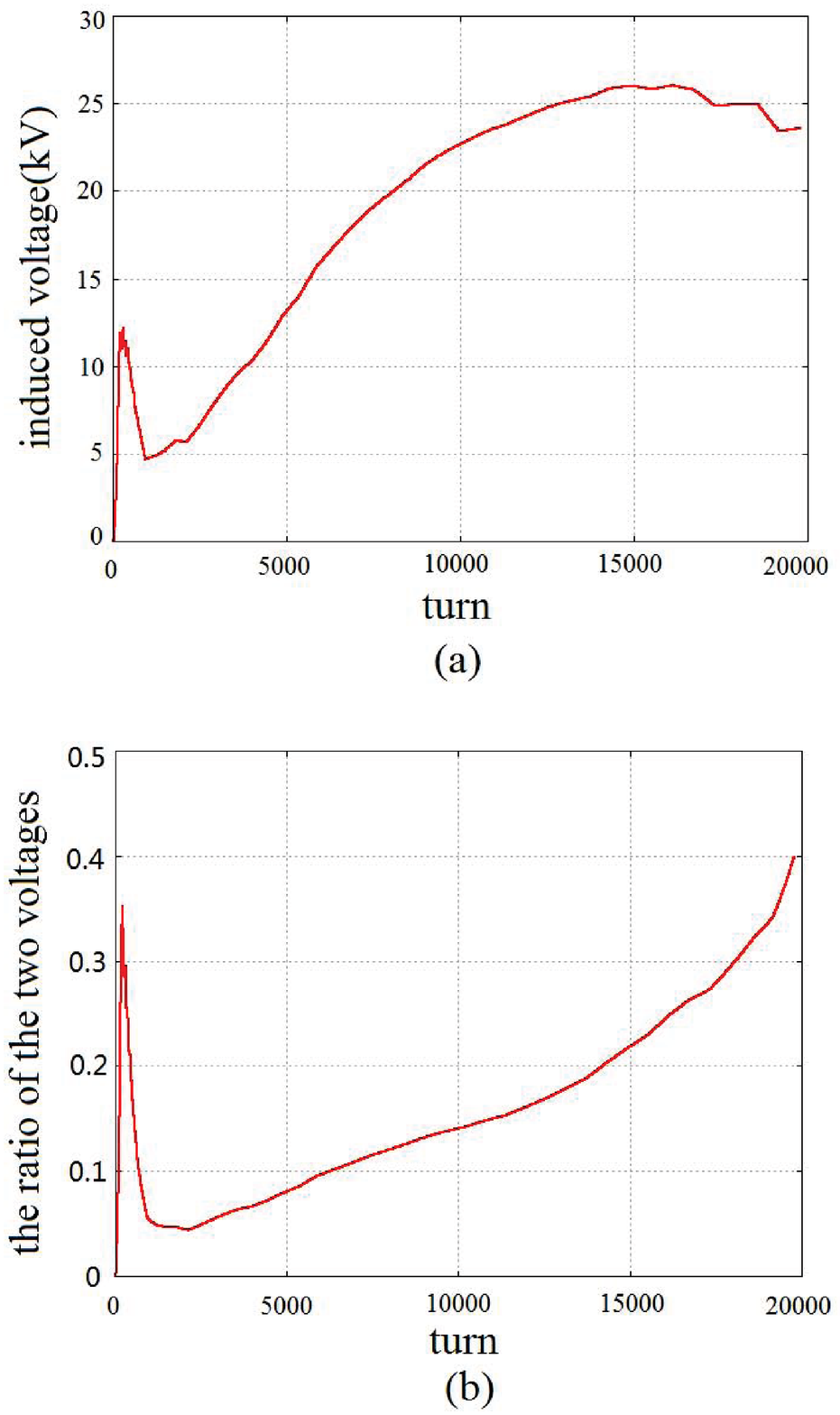}
\figcaption{\label{Fig.8.}   (a) The curve of the beam-induced voltage; (b) The ratio of the beam-induced voltage to the generator voltage in one acceleration period }
\end{center}

\end{multicols}
\ruleup
\begin{center}
\includegraphics[width=14cm]{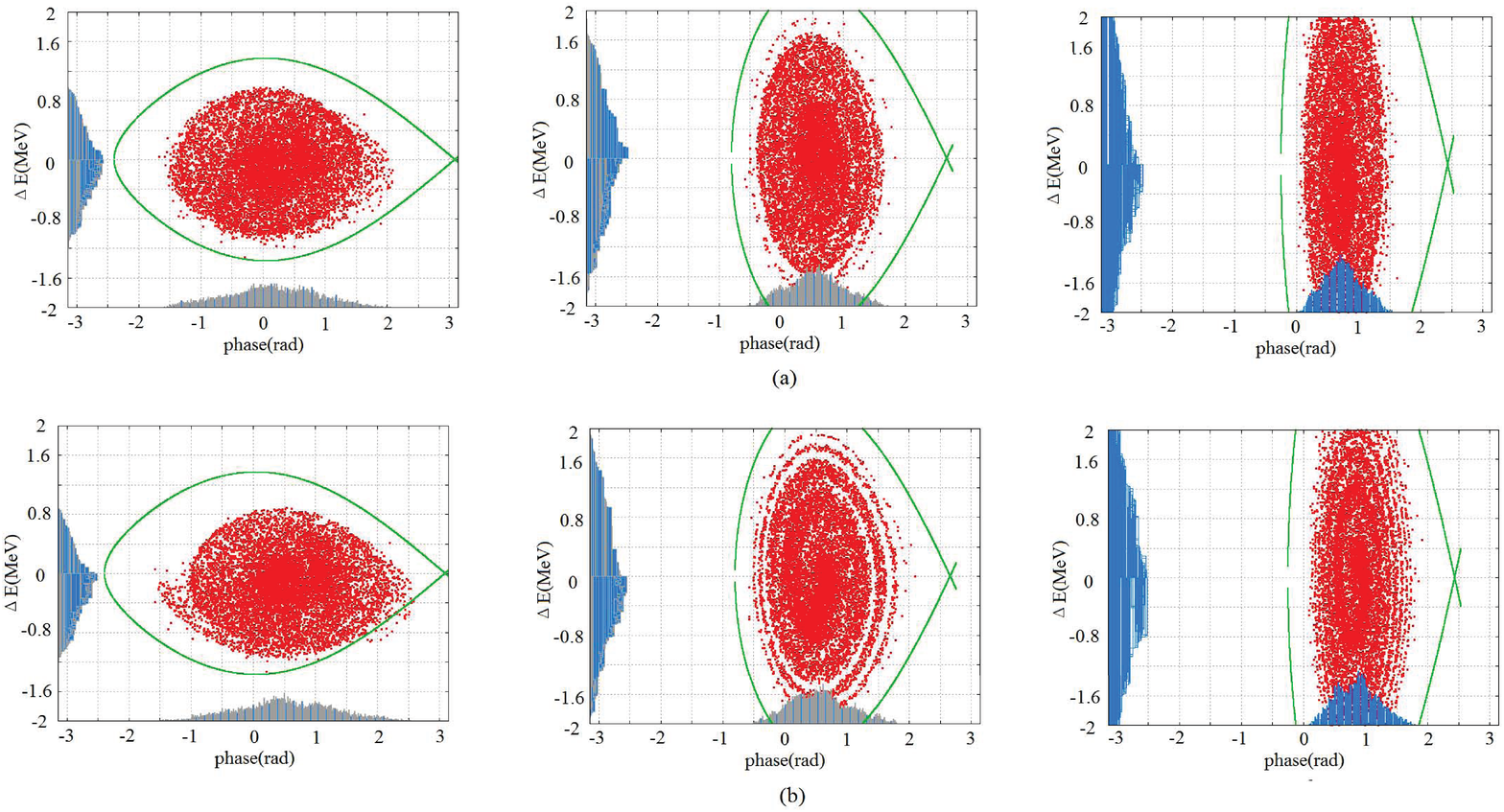}
\figcaption{\label{Fig.9.} The beam distributions in phase space (right) with and (left) without beam loading at (a) 200turn; (b) 2000turn; (c) 5000 turn}
\end{center}
\ruledown
\begin{multicols}{2}

The simulation results demonstrates the amplitude of the effective voltage acted on the beam will decrease and the synchronous phase will increase with the beam loading effect taken into account in the CSNS/RCS. The beam-induced voltage can reach up to 40\% of the generator voltage. Moreover, due to the beam loading effect, the beam loss will occur mainly in the first 300 turns and the ratio of beam loss is nearly 1\%, which suggests that the beam loading compensation is needed in the beam operation.

\section{Summary}

In order to study the beam loading effect in the CSNS/RCS, the physical model of beam loading effect is developed in the simulation code ORIENT, including model of the RLC circuit, the calculation of the current spectrum based on the method of Fast Fourier transform and the beam-induced voltage based on the method of Laplace transform. By using the code, the simulation of beam loading effect has been performed for CSNS/RCS and some results are drawn.

\vspace{5mm}

\acknowledgments{The authors would like to thank HUANG Wei-Ling and LI Xiao for helpful discussions
and measurements.}

\end{multicols}

\vspace{-2mm}
\centerline{\rule{80mm}{0.1pt}}
\vspace{2mm}

\begin{multicols}{2}

\end{multicols}


\end{document}